\documentclass{article}
\usepackage[utf8]{inputenc}
\usepackage[final]{pdfpages}
\usepackage[colorlinks=true]{hyperref}

\usepackage[backend=biber,style=alphabetic,sorting=ynt]{biblatex}

\title{Benchmarking Apache Arrow Flight - A wire-speed protocol for data transfer, querying and microservices}
\author{Tanveer Ahmad, Zaid Al Ars and H. Peter Hofstee}
\date{April 2022}

\begin{document}

\maketitle

\begin{abstract}
  
  Moving structured data between different big data frameworks and/or data warehouses/storage systems often cause significant overhead. Most of the time more than 80\% of the total time spent in accessing data is elapsed in serialization/de-serialization step.       
  Columnar data formats are gaining popularity in both analytics and transactional databases.  
  Apache Arrow, a unified columnar in-memory data format promises to provide efficient data storage, access, manipulation and transport. In addition, with the introduction of the Arrow Flight communication capabilities, which is built on top of gRPC, Arrow enables high performance data transfer over TCP networks. Arrow Flight allows parallel Arrow RecordBatch transfer over networks in a platform and language-independent way, and offers high performance, parallelism and security based on open-source standards.  
  In this paper, we bring together some recently implemented use cases of Arrow Flight with their benchmarking results. These use cases include bulk Arrow data transfer, querying subsystems and Flight as a microservice integration into different frameworks to show the throughput and scalability results of this protocol. 
  We show that Flight is able to achieve up to 6000 MB/s and 4800 MB/s throughput for DoGet() and DoPut() operations respectively. On Mellanox ConnectX-3 or Connect-IB interconnect nodes Flight can utilize upto 95\% of the total available bandwidth. Flight is scalable and can use upto half of the available system cores efficiently for a bidirectional communication. For query systems like Dremio, Flight is order of magnitude faster than ODBC and turbodbc protocols. Arrow Flight based implementation on Dremio performs 20x and 30x better as compared to turbodbc and ODBC connections respectively. We briefly outline some recent Flight based use cases both in big data frameworks like Apache Spark and Dask and remote Arrow data processing tools. We also discuss some limitations and future outlook of Apache Arrow and Arrow Flight as a whole. 
\end{abstract}



\maketitle

\section{Introduction}
Transferring data between databases/data storage systems and client programs in bulk amounts for machine learning applications or statistical analysis, is a common task in data-science. This operation is rather expensive as compared to subsequent operations and becomes even more expensive when the data storage server runs on a different machine or in a cloud environment.
Open-source data science developers/researchers and organizations heavily rely on the Python/R-based data-science eco-system. Apache Parquet, ORC, Avro, and HDFS are commonly used binary formats to store data in compressed form other than text based CSV format. Data serialization and de-serializations on different data processing pipelines (e.g., converting to Pandas Dataframes) built in this eco-system add an additional overhead before actual data can be processed. If data has to be transferred from a remote DBMS server using the DBMS network protocol to these applications, it becomes more expensive due to: i.\ reading from row-oriented DBMS, ii.\ transferring via slower ODBC/JDBC network protocols, iii.\ converting it to required columnar format. So converting row-store data to columnar format is always a major source of inefficiency in data analytics pipelines~\cite{mainliningDBs}. As these formats are designed to store imputable data structures (write-once and read-many), because of this reason they are supposed to be helpful in data analytics workloads only and are not susceptible for transactional workloads. Conventionally, row-store DBMS has been used for OLTP workloads, however recent work by SAP HANA~\cite{SAPHANA} paves the way to bring up column-oriented databases to the mainstream by introducing a highly scalable and efficient query processing engine for both transactional and analytics workloads. TiDB~\cite{tidb} is an open-source example of such a hybrid database system, that supports both transactional and analytical workloads. It is both distributed and MySQL compatible, featuring horizontal scalability, strong consistency, and high availability. 

Apache Arrow provides an open-standard unified in-memory and columnar data format. It alleviates the need of serialization/de-serialization of data through a common format and by providing interfaces for different languages, which makes zero-copy inter-process communication possible. Although Arrow targets mainly OLAP (read-only) workloads, OLTP workloads can still benefit from it. Arrow Flight a submodule in Arrow project provides a protocol to implement a service which can send and receive Arrow (RecordBatches) data streams over the network.   
 

In this work, we discuss the current state-of-the-art for Arrow Flight in terms of development and its applications. We benchmark the performance of Arrow Flight on both the client-server model as well as on the cluster environment and examine the actual speed and bottlenecks in Arrow data transfer, query execution and in microservices usage in different distributed big-data/machine learning frameworks.

The reminder of this paper is organized as follows: In section "\nameref{subsec:Background}", we discuss the Apache Arrow and Arrow Flight internals and architecture in details, followed by a "\nameref{sec:benchmarks}" sections, where Arrow Flight localhost and client-server benchmarks are discussed. We then describe query subsystem and Arrow Flight as microservice integration into some data analytic frameworks in "\nameref{subsec:UseCases}" section. At the end in "\nameref{subsec:Conclusion}" section we conclude this work by outlining some future approaches and use cases.   

\section{Background}
\label{subsec:Background}
In this section, we outline the architectural and design aspects of Apache Arrow and its APIs, particularly Arrow Flight, in detail.
\subsection{Apache Arrow}
Apache Arrow~\cite{ApacheArrow} intends to become a standard columnar format for in-memory data analytics. Introduced in 2015, Arrow provides cross-language interoperability and IPC by supporting different languages, C, C++, C\#, Go, Java, JavaScript, MATLAB, Python, R, Ruby, and Rust. Arrow also provides support for heterogeneous platforms in the form of rapids.ai for GP-GPUs and Fletcher for FPGA systems \cite{fletcher}. Apache Arrow is increasingly extending its eco-system by supporting different APIs (e.g., Parquet, Plasma Object Store, Arrow Compute, etc.) and many open-source libraries/tools are integrating Arrow inside them for efficient data manipulation and transfer. For example, TensorFlow has recently introduced the TensorFlow I/O~\cite{tensorio} module to support the Arrow data format, the Dremio big data framework is built around the Apache Arrow eco-system, pg2arrow (a utility to query PostgreSQL relational database), turbodbc which supports queries in Arrow format, etc.

Arrow stores data in contiguous memory locations to make the most efficient use of CPU's cache and vector (SIMD) operations. In the Arrow format, data entries (records) are stored in a table called a RecordBatch. An example of a RecordBatch with three records (rows) and three fields (columns) is shown in Table~\ref{tab:rb}. As shown in Table~\ref{tab:buffers}, each field in the RecordBatch table is stored in a separate memory region in a manner that is as contiguous as possible in memory. This memory region is called an Arrow Field or Array which can store data of different types---i.e., int, float, UTF8 characters, binary, timestamps, lists and nested types. Depending on the data types, fields can have multiple Arrow Buffers to store extra information about the data, such as a validity bit for nullable data types, or offsets in the case of variable-sized lists. Through this approach, accessing data from random locations and in parallel with a minimum amount of pointers traversing becomes possible. 
This approach makes Arrow less efficient particularly in large write-updates of variable length strings which is a point of concern for using Arrow in transactional workloads. 


\begin{table}[]
\centering
  \caption{Example table stored as an Arrow RecordBatch}
  \label{tab:rb}
\begin{tabular}{lll}
\hline
X     & Y       & Z      \\ \hline
555   & "Arrow" & 5.7866 \\
56565 & "Data"  & 0.0      \\
null  & "!"     & 3.14   \\ \hline
\end{tabular}
\end{table}

\begin{table}[]
\centering
  \caption{Arrow Buffers layout for data in Table \ref{tab:rb}}
  \label{tab:buffers}
\begin{tabular}{llllll}
\hline      & \multicolumn{5}{c}{Arrow Buffers for:} \\ \hline
      & \multicolumn{2}{c}{Field X} & \multicolumn{2}{c}{Field Y}& Field Z \\ \hline
Index & \begin{tabular}[c]{@{}l@{}}Validity\\ (bit)\end{tabular} & \begin{tabular}[c]{@{}l@{}}Values\\ (Int32)\end{tabular} & \begin{tabular}[c]{@{}l@{}}Offsets\\ (Int32)\end{tabular} & \begin{tabular}[c]{@{}l@{}}Values\\ (Utf8)\end{tabular} & \begin{tabular}[c]{@{}l@{}}Values\\ (Double)\end{tabular} \\ \hline
0     & 1                                                        & 555                                                      & 0                                                         & A                                             & 5.7866                                                    \\
1     & 1                                                        & 56565                                                    & 5                                                         & r                                             & 0.0                                                         \\
2     & 0                                                        & null                                                     & 9                                                         & r                                             & 3.14                                                      \\
3     &                                                         &                                                      &                                                           & o                                             &                                                           \\
4     &                                                        &                                                      &                                                           & w                                             &                                                           \\
5     &                                                          &                                                          &                                                           & D                                             &                                                           \\
6     &                                                          &                                                          &                                                           & a                                             &                                                           \\
7     &                                                          &                                                          &                                                           & t                                             &                                                           \\
8     &                                                          &                                                          &                                                           & a                                             &                                                           \\
9     &                                                          &                                                          &                                                           & !                                             &                                                           \\ \hline
\end{tabular}
\end{table}

\begin{table}[]
\centering
  \caption{Schema for RecordBatch in Table \ref{tab:rb}}
  \label{tab:schema}
\begin{tabular}{|l|l}
\cline{1-1}
Field X: Int32 (nullable), &  \\
Field Y: Utf8,             &  \\
Field Z: Double            &  \\ \cline{1-1}
\end{tabular}
\end{table}
Each RecordBatch contains metadata, called a schema, that represents the data types and names of stored fields in the RecordBatch. Table~\ref{tab:schema} shows the schema of the example Arrow RecordBatch shown in Table~\ref{tab:rb}.

\begin{figure*}[ht]
  \centering
  \includegraphics[width=\linewidth]{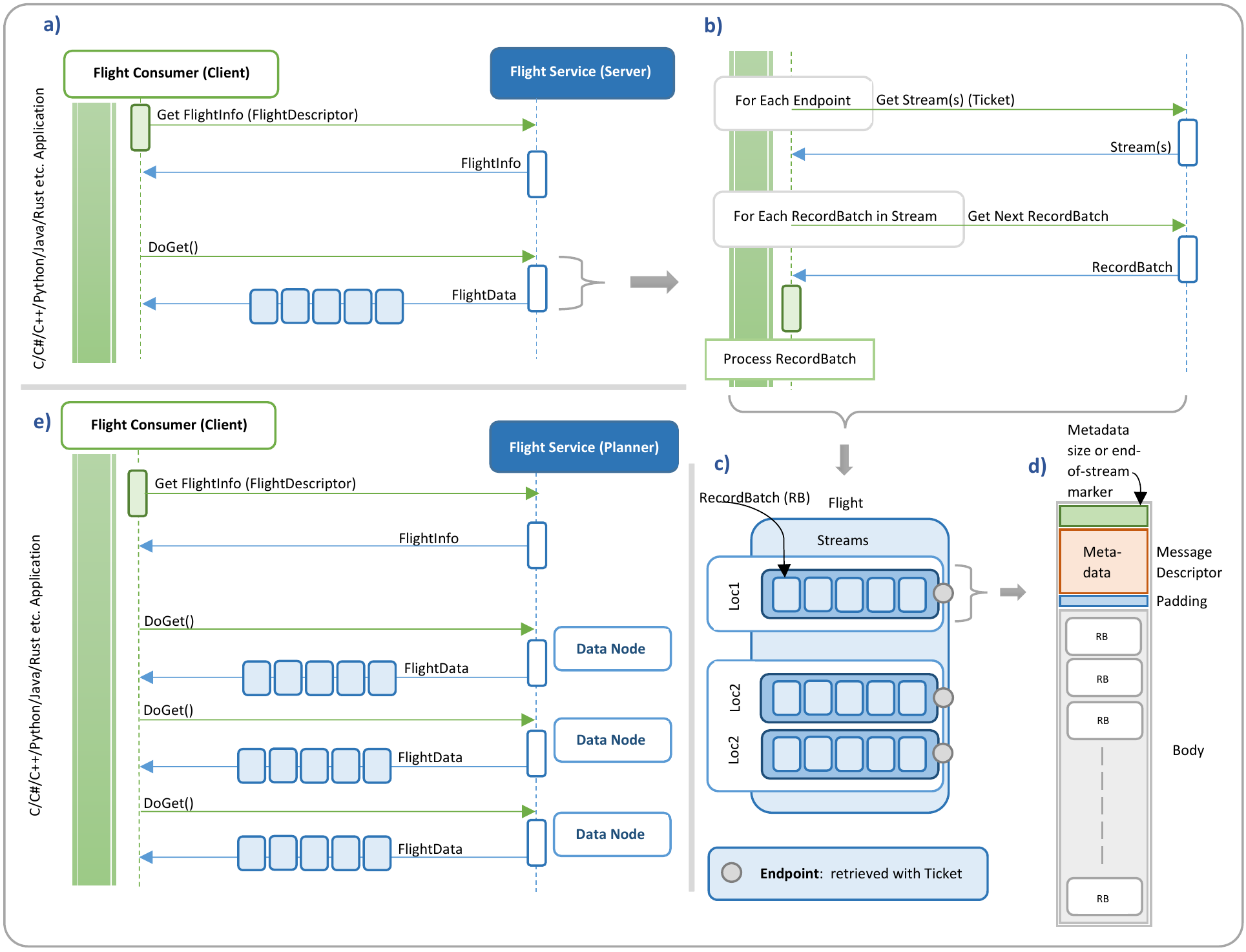}
  \caption{a) Arrow Flight client-server communication protocol dataflow diagram for DoGet() operation. b) In depth Flight streams communication dataflow for accessing RecordBatches in a stream. c) Flight streams endpoint. d) Inter-process communication format. e) Arrow Flight cluster communication protocol dataflow diagram with multiple nodes and a single planner node for DoGet() operation.}
  \label{fig:flight-proto}
\end{figure*}
\subsection{Arrow Flight}
Arrow Flight~\cite{ApacheArrowFlight} provides a high performance, secure, parallel and cross-platform language support (using the Apache Arrow data format) for bulk data transfers particularly for analytics workloads across geographically distributed networks. Using Apache Arrow as standard data format across all languages/frameworks as well as on the wire, Arrow Flight data (Arrow RecordBatches) does not require any serialization/de-serialization when it crosses process boundaries. As Arrow Flight operates directly on Arrow RecordBatches without accessing data of individual rows as compared to traditional ODBC/JDBC interfaces, it is able to provide high performance bulk operations. Arrow Flight supports encryption out of the box using gRPC’s built in TLS/OpenSSL capabilities. Simple user/password authentication scheme is provided out-of-the-box in Arrow Flight and provides extensible authentication handlers for some advanced authentication schemes like Kerberos. 

In basic Arrow Flight communication, a client initiates the communication by sending the \verb|GetFlightInfo()| command to the server. In case of a successful connection, the server replies with available Flights by sending back \verb|FlightInfo| information, which contains so-called \verb|Ticket|s that define locations (or \verb|Endpoint|s) of streams of RecordBatches at the server side. Then, the \verb|DoPut()| command is used by the client to send a stream of RecordBatches to the server, and the \verb|DoGet()| command is used by the server to send a stream back to the client. Both these commands are initiated by the client. Figure~\ref{fig:flight-proto}(a) shows the data flow protocol diagram for an example Flight communication with the \verb|GetFlightInfo()| and \verb|DoGet()| commands. In Figure~\ref{fig:flight-proto}(b), the client uses the \verb|GetStream| command to request one or more streams of RecordBatches by calling their \verb|Ticket| information. Figure~\ref{fig:flight-proto}(c) shows the internal structure of Flight communication. Figure~\ref{fig:flight-proto}(d) shows the Flight protocol description within each stream, which contains the stream metadata and RecordBatches.
\begin{figure*}[t]
  \includegraphics[width=\linewidth]{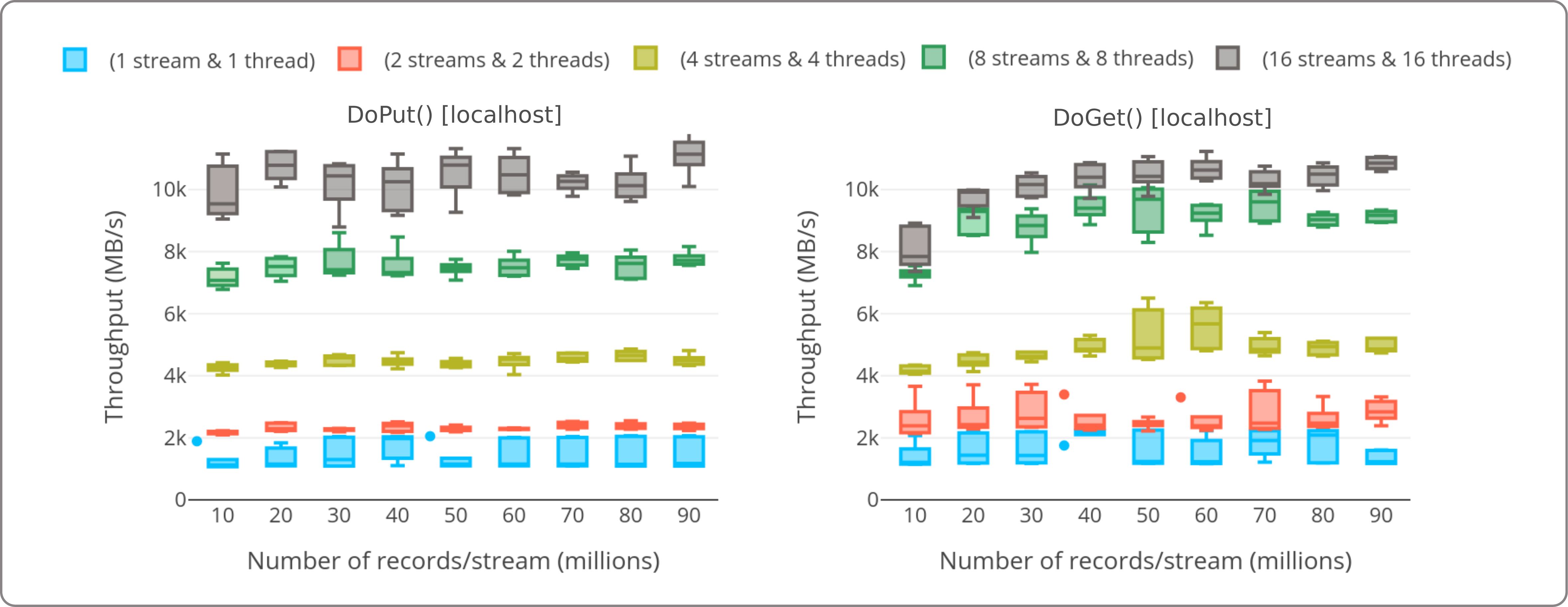}
  \caption{Arrow Flight DoPut() and DoGet() throughput with multiple stream/threads with varying number of records per stream (10-90 million) on a localhost.}
  \label{fig:flight-local}
\end{figure*}
\begin{figure*}[t]
  \includegraphics[width=\linewidth]{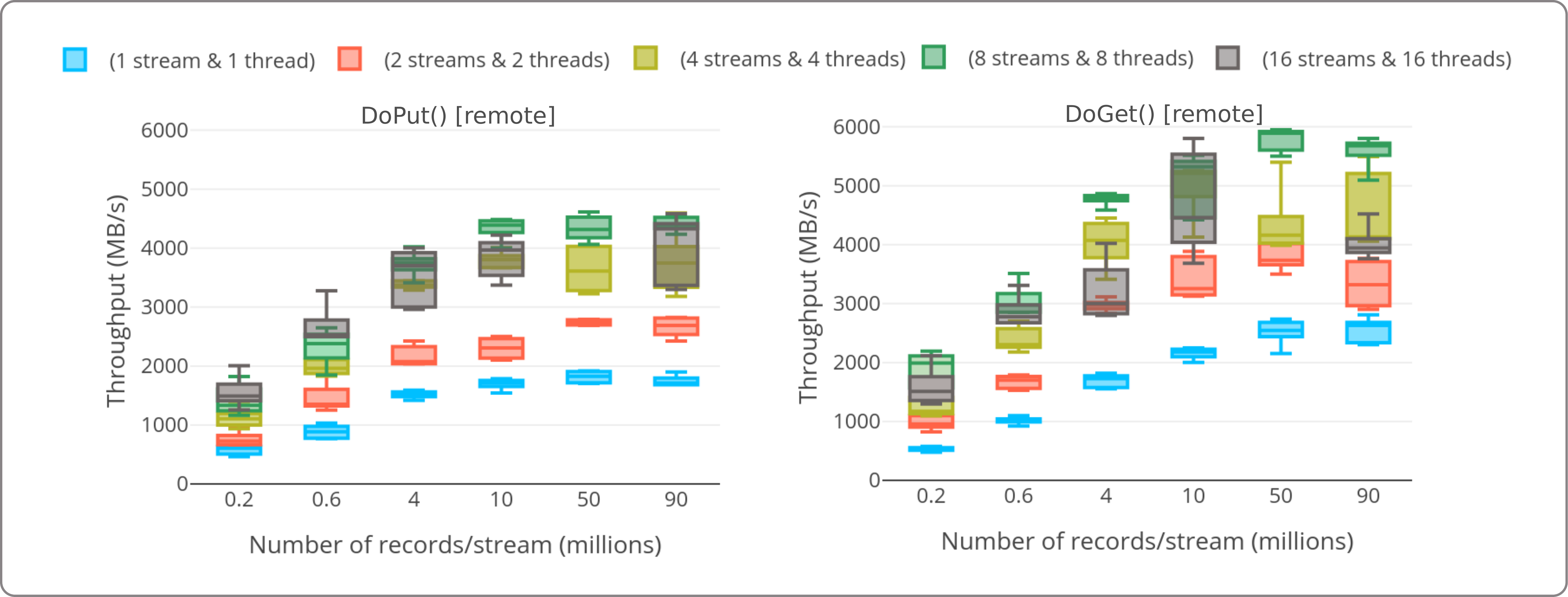}
  \caption{Arrow Flight DoPut() and DoGet() throughput with multiple stream/threads with varying number of records per stream (0.2-90 million) on a remote client-server nodes connected through a Mellanox ConnectX-3 or Connect-IB InfiniBand adapter.}
  \label{fig:flight-remote}
\end{figure*}
    
    Flight services can handle multiple Flight connections in a cluster environment and can differentiate between them using a Flight descriptor, which can define the composition of Flight connections with batch size, and either file name or SQL query command as shown in Figure~\ref{fig:flight-proto}(e).    







\subsection{Distributed Columnar-store and Analytics}
Relational databases are optimized for transactional workloads, which makes them less efficient to support the needs of modern analytics applications. Querying billions of rows on demand from a row-oriented database for analytics and statistical purposes becomes a bottleneck in real-time column-oriented data analytics workloads. Moreover, production-ready analytics workloads and ML pipelines mostly use public clusters and cloud computing infrastructures for a couple of reasons including security, scalability, high-availability, lower costs and on-demand easy to deploy functionality. All major cloud service providers present their own distributed datastores like Google (Cloud SQL for OLTP and  BigQuery for OLAP systems), Amazon (AWS Redshift and AWS S3) and Azure (Cosmos DB and Synapse Analytics) for both analytics and SQL. Apache Hive, Dremio, Presto and Apache Impala are a couple of BI/data science SQL based engines built to communicate with distributed datasets using different storage formats. The support of universal storage formats (like HDFS, ORC, CSV, Parquet) makes these systems flexible to export data in any form and to any system for further processing. For such a distributed data-store environment, it essential to provide high-throughput methods to communicate large datasets between systems. The Arrow format also supports local, Parquet, HDFS and S3 distributed file systems, which makes Arrow Flight an important differentiator for Arrow-based applications. 

\subsection{DB-X Data Export to External tool}
\begin{figure}[ht]
  \centering
  \includegraphics[width=\linewidth]{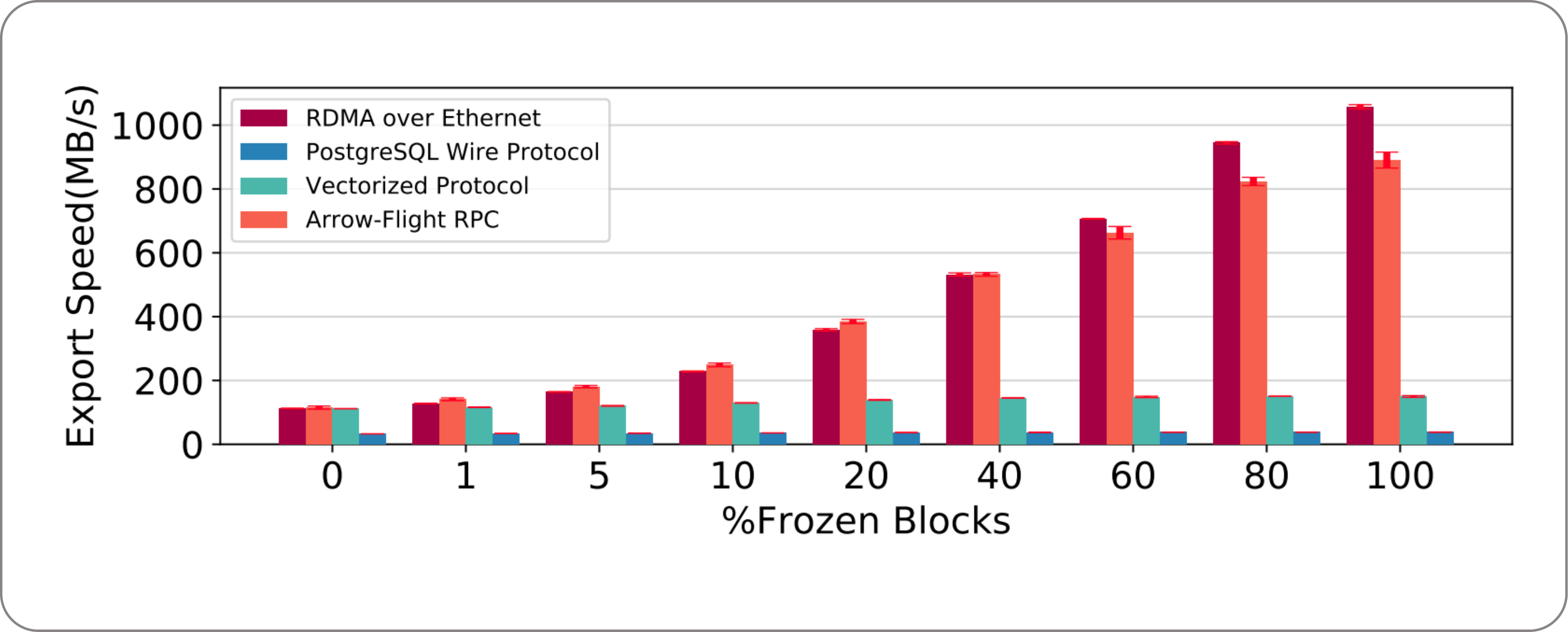}
  \caption{Data Export – Measurements of export speed with different export mechanisms in DB-X, varying \% of hot blocks~\cite{mainliningDBs}.}
  \label{fig:mainliningDBX}
\end{figure}

As reported in~\cite{mainliningDBs}, the authors evaluate the data export (to an external tool) performance of DB-X. They compared four different export methods, (1) client-side RDMA, (2)
Arrow Flight RPC, (3) vectorized wire protocol from~\cite{DataHostage},
and (4) row-based PostgreSQL wire protocol. They used the TPC-C \verb|ORDER_LINE| table with 6000 blocks (approximately 7 GB total size) on the server. By varying the \% of frozen blocks in DB-X they study the impact of concurrent transactions on export speeds. Figure~\ref{fig:mainliningDBX} shows when all the blocks are frozen, RDMA saturates the bandwidth while Flight uses up to 80\% of the available bandwidth. When the system has to materialize every block, the performance of Arrow Flight drops to be equivalent to the vectorized wire protocol. RDMA performs slightly worse than Arrow Flight with a large number of hot blocks, because Flight has the materialized block in its CPU cache, whereas the NIC bypasses this cache when sending data. Overall the experiment shows that (de)-serialization is the main bottleneck in achieving better data export speeds in DBMS. Using a common data format in DBMS like Arrow can boost the export speeds in-conjunction with the Arrow Flight wire protocol. 
\section{Data Transfer Benchmarks}
\label{sec:benchmarks}
Bulk data transfers over long-haul networks has become an integral part of modern data science applications. Reading and extracting required data from remote datasets through remote data services like ODBC or JDBC is inefficient and lacks support for current applications and frameworks. Although in the past decade, file-based (text formats like CSV, JSON and binary formats like Avro, ORC and Parquet) data warehousing has become popular, still raw data needs serialization/de-serialization to a particular format when accessed/used by different applications on remote/local servers. With Arrow Flight, a unified Arrow columnar data format can be used, which provides both over-the-wire data representation as well as a public API for different languages and frameworks. This in turn eliminates much of the serializations overheads associated with data transport. 

\subsection{Evaluation system}
Arrow Flight based bulk data transfer benchmarks in this section 
are executed on the SurfSara Cartesius~\cite{SurfSara} HPC cluster (part of the Dutch national supercomputing infrastructure). Each CPU-only node is equipped with a dual socket Intel Xeon Processor (E5-4650) running at 2.7 GHz. Each processor has 32 physical cores with support of 64 hyper-threading jobs. A total of 256-GBytes of DDR4 DRAM with a maximum of 59.7 GB/s bandwidth is available for each node. All nodes are connected through Mellanox ConnectX-3 or Connect-IB InfiniBand adapter providing 4×FDR (Fourteen DataRate) resulting in 56 Gbit/s inter-node bandwidth.

\subsection{Client-Server Microbenchmarks}
To measure the absolute speed and performance of Arrow Flight, we use the Flight built-in performance benchmark written in C++ in both localhost and on a network in a client-server setting.
In localhost, a loopback network interface is established on a single computer node. Usually, in client-server model server controls the communication between associated client(s) over the network. Figure~\ref{fig:flight-local} shows throughput variation of Arrow Flight data transport for localhost while Figure~\ref{fig:flight-remote} shows throughput for client-server settings. We use 1, 2, 4, 8 and 16 streams in parallel with each stream having 10-90 million records. Each record contains 32 bytes. On localhost, both \verb|DoPut()| and \verb|DoGet()| functions give a throughput in the order of 1GB/s for single stream up to 10GB/s with 16 streams in parallel. As the localhost processor has 16 physical cores on two sockets with support of 32 hyper-threading jobs. So Arrow Flight performance shows a significant increase in throughput when more parallel streams are employed. We also observe that increasing the parallel streams more than 16 decreases the overall performance. We run the client-server benchmark in a network~\cite{SurfSara} in which every node has a Mellanox ConnectX-3 or Connect-IB (Haswell thin nodes) InfiniBand adapter providing 4 × FDR (Fourteen Data Rate) resulting in 56 Gbit/s inter-node bandwidth. We see the same trend in this remote data transfer setting with throughput increasing from 1.2GB/s to 1.65GB/s for \verb|DoPut()| while \verb|DoGet()| achieves 1.5GB/s to 2GB/s throughput with up to 16 streams in parallel. 
\begin{figure}[ht]
  \centering
  \includegraphics[width=\linewidth]{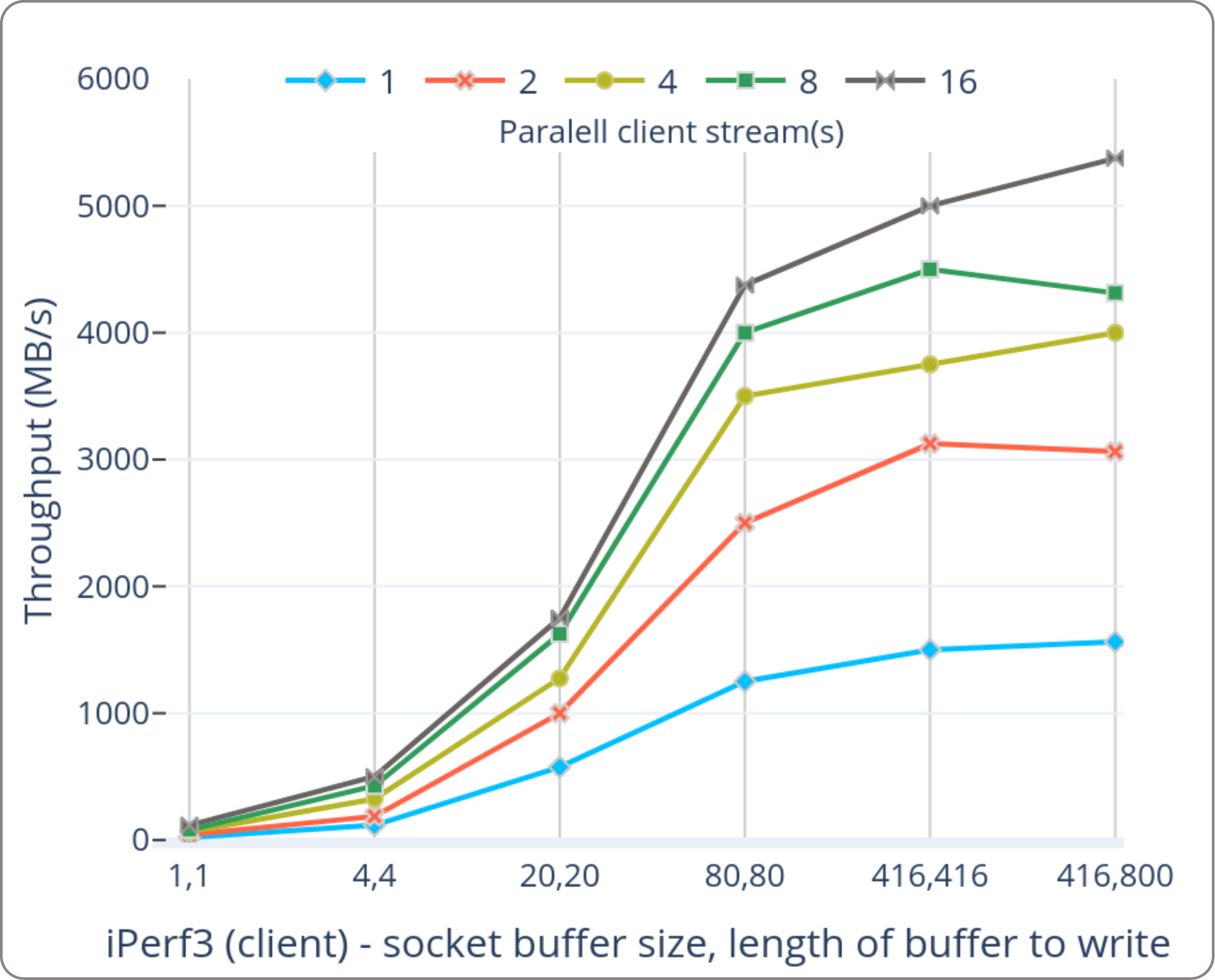}
  \caption{iPerf3 based client/server benchmarking of TCP data send/receive overall transfer throughput on a Mellanox ConnectX-3 or Connect-IB InfiniBand adapter based system.}
  \label{fig:tcp}
\end{figure}
\begin{figure}[ht]
  \centering
  \includegraphics[width=\linewidth]{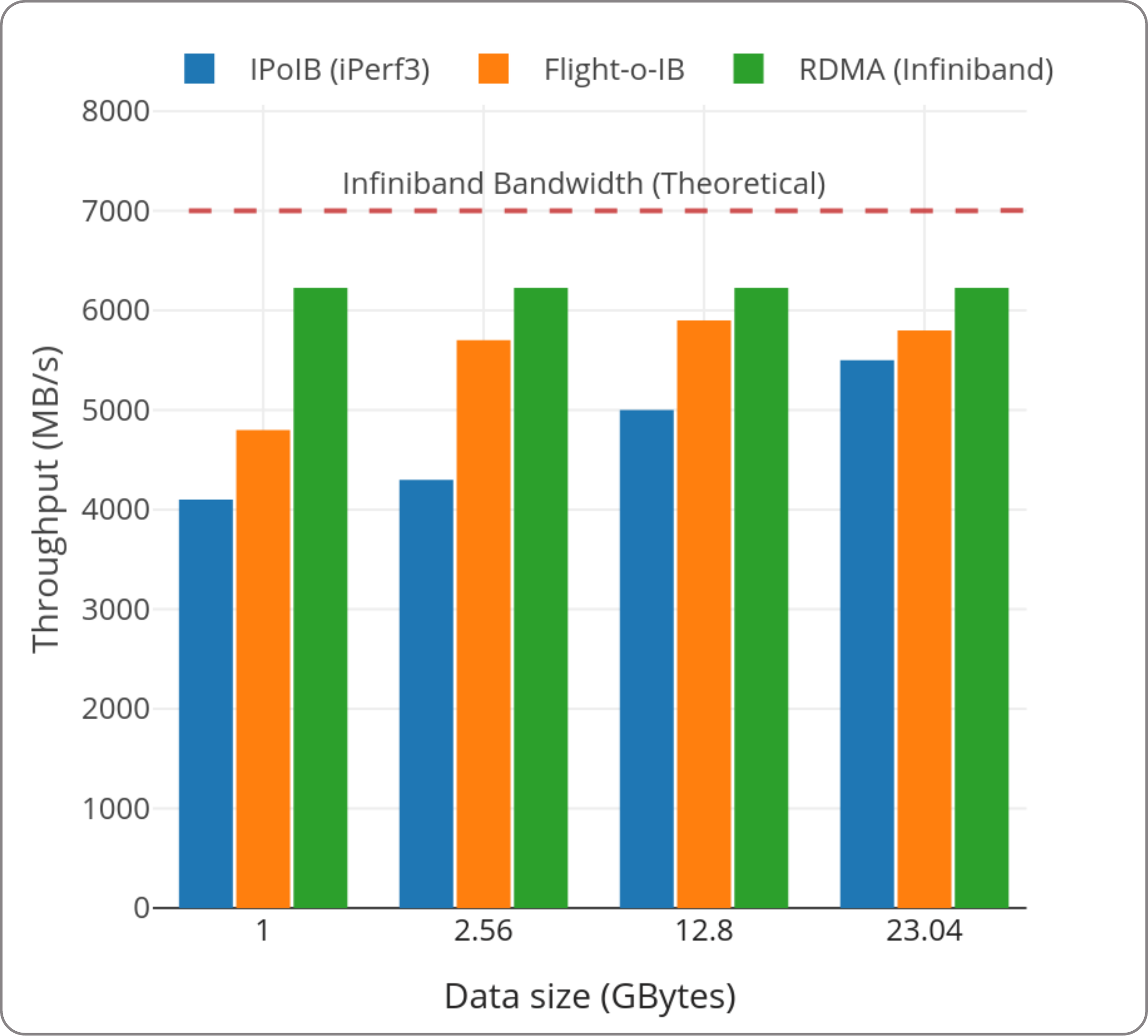}
  \caption{Throughput comparison of IPoIB using iPerf3,  Flight-over-IB and RDMA (Infinaband) on a Connect-IB InfiniBand adapter based client-server remote system.}
  \label{fig:rdma}
\end{figure}
To compare the throughput of Arrow Flight with other common communication protocols, Figure~\ref{fig:rdma} measures the throughput of Flight over InfiniBand (Flight-o-IB) and two other communication protocols on the same network for remote client-server communication: 1.~the TCP protocol over InfiniBand (TCP-o-IB), commonly used for long-haul data communication, and 2.~RDMA over InfiniBand (RDMA-o-IB) protocol, commonly used for high-throughput cluster-based communication. To measure TCP throughput, we use the iPerf3~\cite{iperf3} network performance measurement tool with multiple parallel streams, which is  able to measures raw TCP throughput as shown in Figure~\ref{fig:tcp} with minimal overhead.  
For RDMA throughput, we use the \verb|ib_write_bw| (InfiniBand write bandwidth) tool which is part of the Perftest Package~\cite{mellanox}. Figure~\ref{fig:rdma} shows that RDMA is able to achieve a high throughput of 6.2GB/s (close to the theoretical max of 7GB/s) for a wide range of data sizes. TCP, on the other hand, has a low throughput of about 2GB/s for small data sizes (256B) that increases slowly as the data size increases, consistently suffering from high overhead for a wide range of data sizes. In contrast, Flight has extremely low bandwidth for very small data sizes of up to 1KB, but then consistently outperforms TCP for larger sizes and is able to achieve about 95\% of the RDMA bandwidth (or more than 80\% of the maximum achievable bandwidth) for data sizes of 2.6GB or larger. This shows the capabilities of Flights to ensure high throughput for bulk data transfers that is comparable to high-throughput protocols such as RDMA over InfiniBand, while retaining the benefits of ease of programmability, security, and allowing access to a wide range of web-based services. 

In addition, the figures show that Flight allows improving the throughput by increasing the number of parallel streams. However, this is not the case for TCP, as increasing the number of streams results in more network congestion and a slight reduction in throughout.

\section{Use Cases}
\label{subsec:UseCases}
This section presents some common use cases of Arrow Flight related to query data transfer from a remote data service, and to Arrow Flight usage in big data frameworks for distributed data transfers in a cluster to be consumed in microservices.  
\begin{figure}[ht]
  \centering
  \includegraphics[width=11cm]{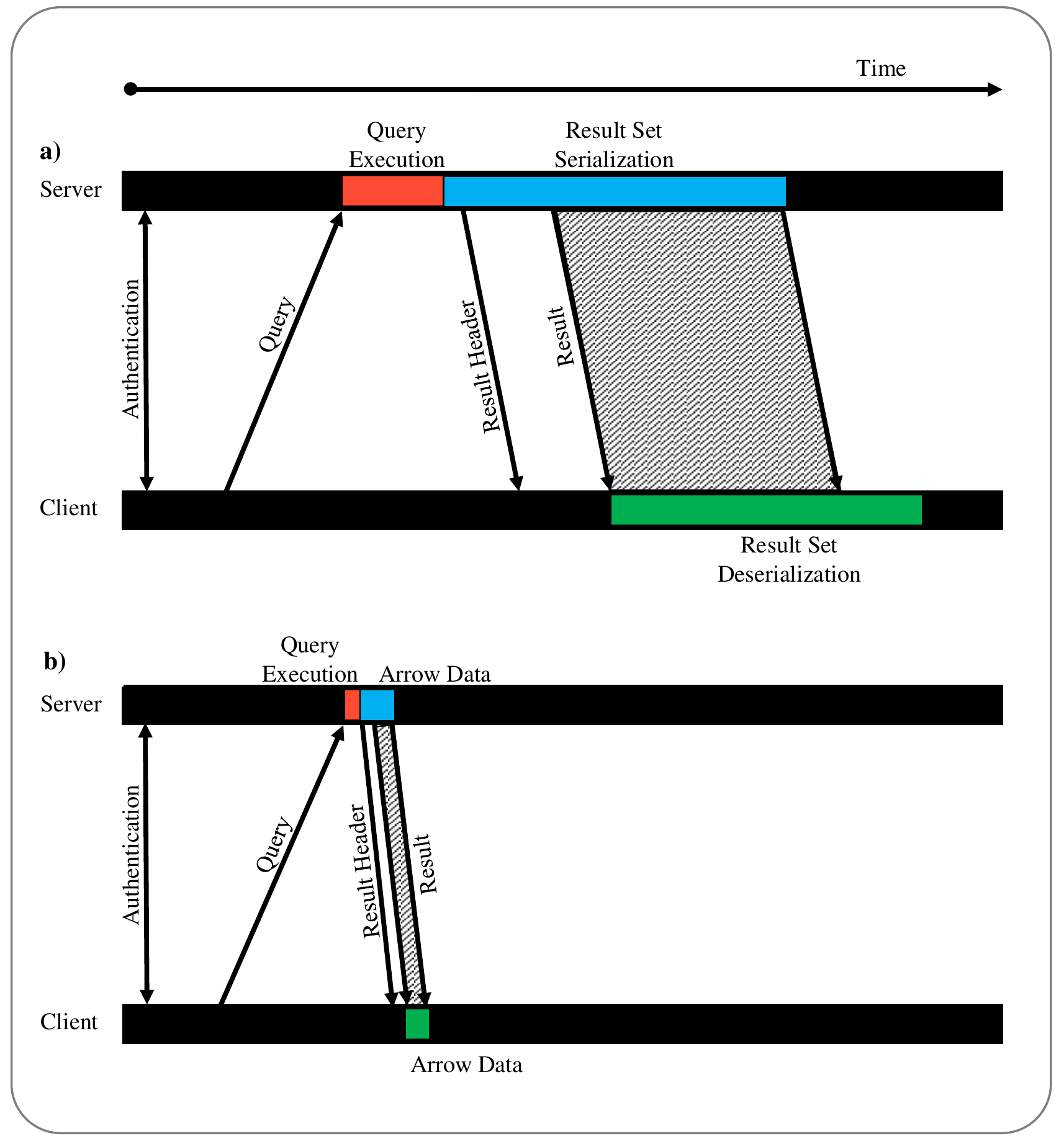}
  \caption{A typical client-server communication for query execution~\cite{DataHostage}. (a) Without Flight: a large amount of time spent in (de)-serialization of the result set is shown. (b) With Flight: the total time spent in query execution on Arrow data with Arrow Flight based communication eliminates any (de)-serialization overhead.}
  \label{fig:DataHostage}
\end{figure}

\begin{figure}[ht]
  \centering
  \includegraphics[width=\linewidth]{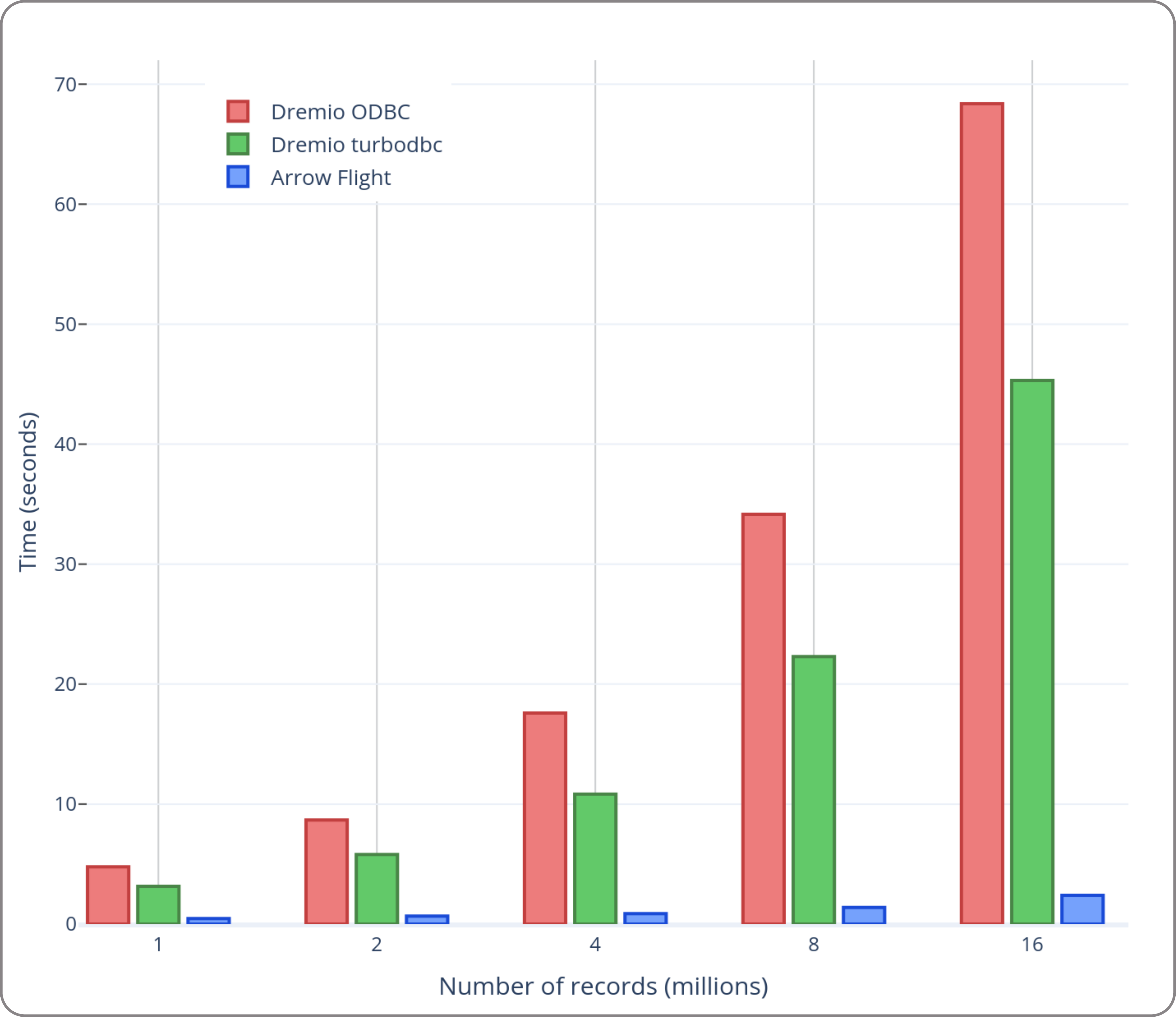}
  \caption{Total time spent in querying NYC Taxi dataset on a remote Dremio client-server nodes with varying number of records (1-16 millions) through ODBC, turbodbc and Flight connections.}
  \label{fig:dremio}
\end{figure}

\begin{figure}[ht]
  \centering
  \includegraphics[width=\linewidth]{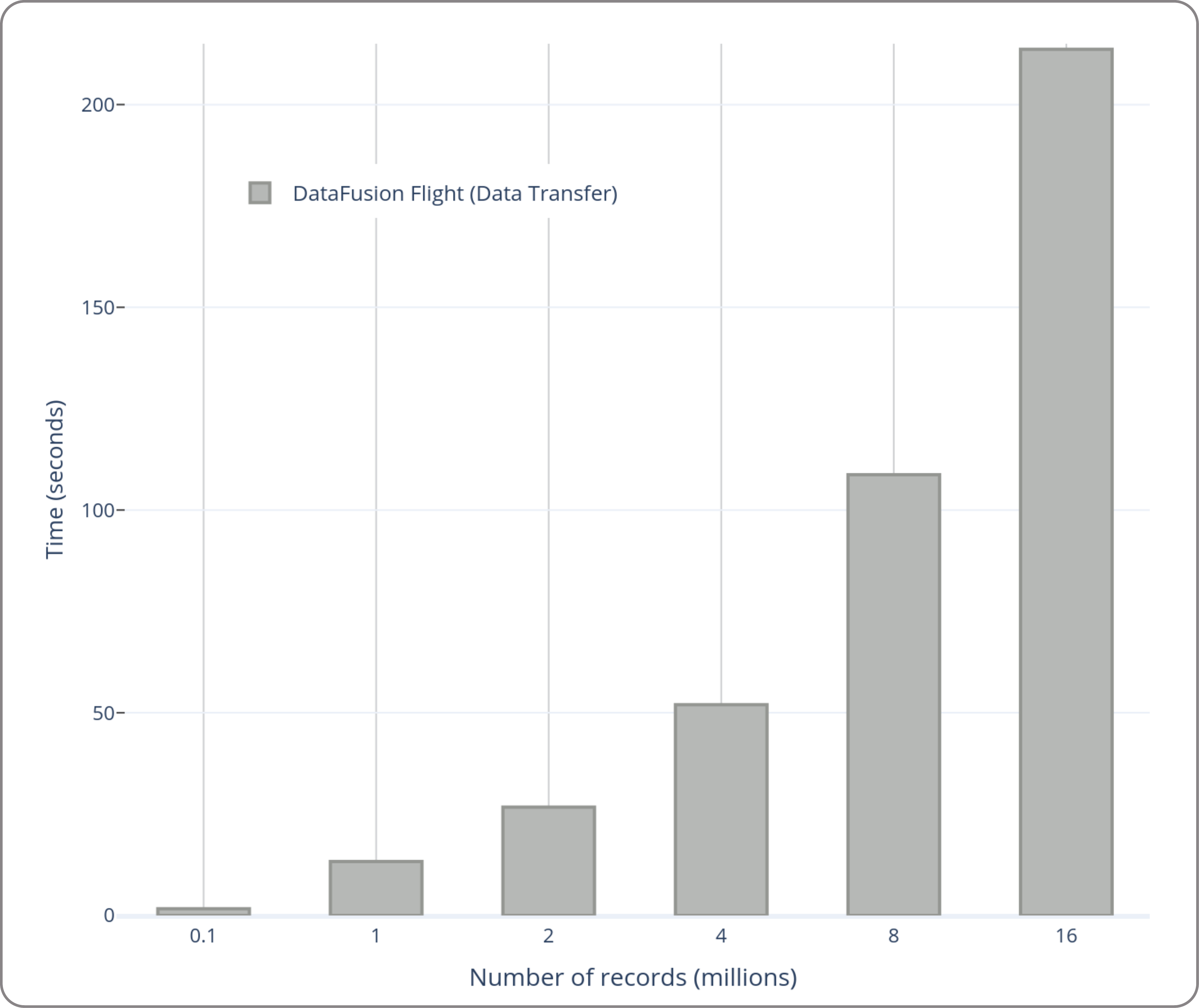}
  \caption{Total time spent in querying NYC Taxi dataset on a remote Arrow Flight based client-server nodes implemented in Data-fusion project with varying number of records (0.1-16 millions).}
  \label{fig:fusion}
\end{figure}

\subsection{Query Subsystem}
Transferring big amounts of data from a local/remote server to a querying client is a common task in both analytics (statistical analysis or machine learning applications) and transactional database systems. As described in~\cite{DataHostage}, a typical client-server data communication scenario is shown in Figure~\ref{fig:DataHostage}, where the communication time is heavily influenced by data serialization overhead.       

Arrow Flight 
provides a standard in-memory unified columnar data format. Exporting Arrow tables to any client language interface avoids (de)-serialization overhead. Better columnar compression techniques and parallel Arrow streams transfer make Arrow Flight ideal for efficient big data transfers.

In this section, we focus on Dremio, an production grade Arrow native analytics framework. We measure the performance metrics of different client side protocols (ODBC and turbodbc) for querying on the Dremio remote client-server and compare the results with Arrow Flight as shown in Figure~\ref{fig:dremio}. We also look at two systems under development: Arrow Datafusion (an Arrow Flight based client-server query API), and FlightSQL (a native SQL API for Arrow data).

\subsubsection*{Dremio - ODBC}
Dremio provides a custom ODBC driver for different client types. We used a Linux based Dremio ODBC driver and used it with the pyodbc Python API to query the NYC Taxi database (in parquet format) from a Dremio server running remotely in a cluster. 
\subsubsection*{Dremio - turbodbc}
We also used the Dremio ODBC driver to connect with a Dremio client through the turbodbc Python API. Here too, we queried the NYC Taxi database (in parquet format) from a Dremio server running remotely in the same cluster. 
\subsubsection*{Dremio - Flight}
Dremio offers client and server Flight endpoint support for Arrow Flight connections that is also authentication OAuth2.0 compliant. Moreover this implementation provides TLS encryption to establish an encrypted connection.

The runtime performance comparison results for all three methods on a single select query are shown in Figure~\ref{fig:dremio}. Arrow Flight based implementation on Dremio performs 20x and 30x better as compared to turbodbc and ODBC connections respectively.    
\subsubsection*{Data-Fusion - Flight}
DataFusion is an in-memory, Arrow-native query engine implemented in Rust. Though this framework is in its initial phases of development, it does support SQL queries against iterators of RecordBatch in both CSV and Parquet file formats. Both the Arrow Flight client and server implementations are available. Figure~\ref{fig:fusion} shows the results we obtained by running the Arrow Flight client-server benchmark provided in Data-Fusion repository. We converted NYC Taxi dataset used in previous Dremio demo to Parquet format and query the same dataset for specific elements in each iteration.     

\subsubsection*{Apache Arrow - FlightSQL}
FlightSQL is a new proposal being implemented by the Apache Arrow community to become a standard way of accessing Arrow data via SQL-like semantics over Flight. The main idea of this framework is to use ODBC and JDBC data access best practices while maintaining the high throughput facilitated by Arrow Flight. 

\subsection{Microservices Integration}
Arrow Flight can be integrated into data transfer and remote data analytics microservices for efficient and parallel processing of Arrow columnar data using many different frameworks like Dask and Spark, as discussed next.

\subsubsection{Flight Data Microservice - Apache Spark~\cite{Spark-source-for-Flight}}
\begin{figure}[ht]
  \centering
  \includegraphics[width=\linewidth]{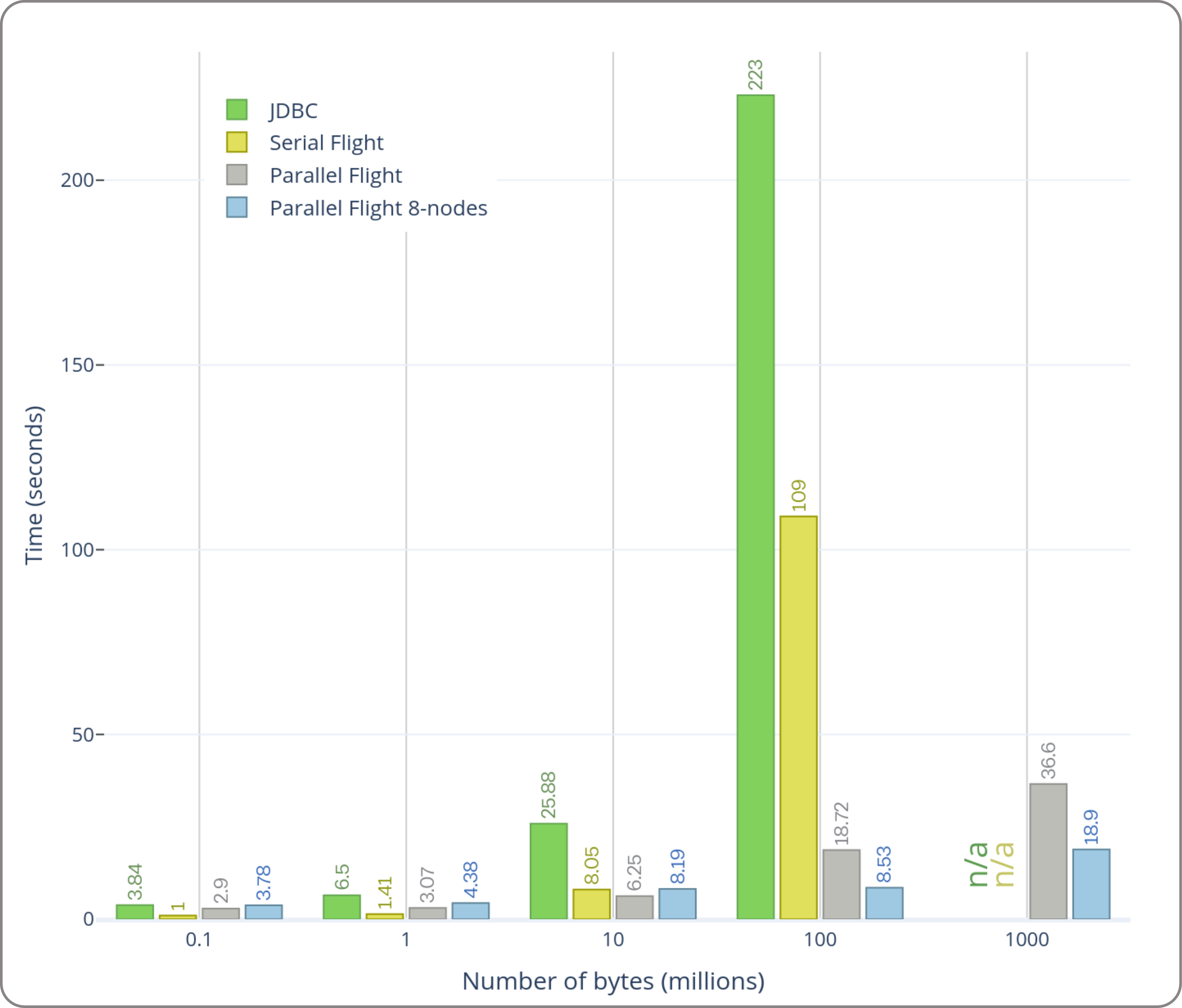}
  \caption{Performance results of Apache Arrow Flight endpoints integration in Apache Spark using the Spark Datasource V2 interface. The results show the total time spent in Spark default JDBC, serial Flight, parallel Flight and parallel Flight with 8-node connections.}
  \label{fig:sparkquery}
\end{figure}

This Arrow Flight based microservice implementation is an early prototype test to showcase the reading of columnar Arrow data, reading in parallel many Flight endpoints as Spark partitions, this design uses the Spark Datasource V2 API to connect to Flight endpoints. Figure~\ref{fig:sparkquery} shows performance comparisons in terms of total time for default JDBC, serial flight, parallel flight and parallel flight with 8 nodes. This test returns $n$ rows to Spark executors and then performs a non-trivial calculation on them. This test was performed on a 4x node EMR with querying a 4x node Dremio AWS Edition (m5d.8xlarge) by the developer. 

\subsubsection{Flight Data Microservice - Apache Spark/TensorFlow Clients~\cite{Flight-SparkandTensorFlowclients}}
A basic Apache Arrow Flight data service with Apache Spark and TensorFlow clients has been demonstrated. In this demo a simple data producer with an InMemoryStore allows clients to put/get Arrow streams to an in-memory store. Existing PySpark DataFrame partitions are mapped by a Spark client to produce an Arrow stream of each partition which are put under the FlightDescriptor. A PyArrow client reads these streams and convert them into Pandas Dataframes. Similarly, a TensorFlow client reads each Arrow stream, one at a time, into an ArrowStreamDataset so records can be iterated over as Tensors~\cite{Flight-SparkandTensorFlowclients}. 
\subsubsection{XGBatch - Pandas/Dask~\cite{xgbatch}}
\begin{figure}[ht]
  \centering
  \includegraphics[width=\linewidth]{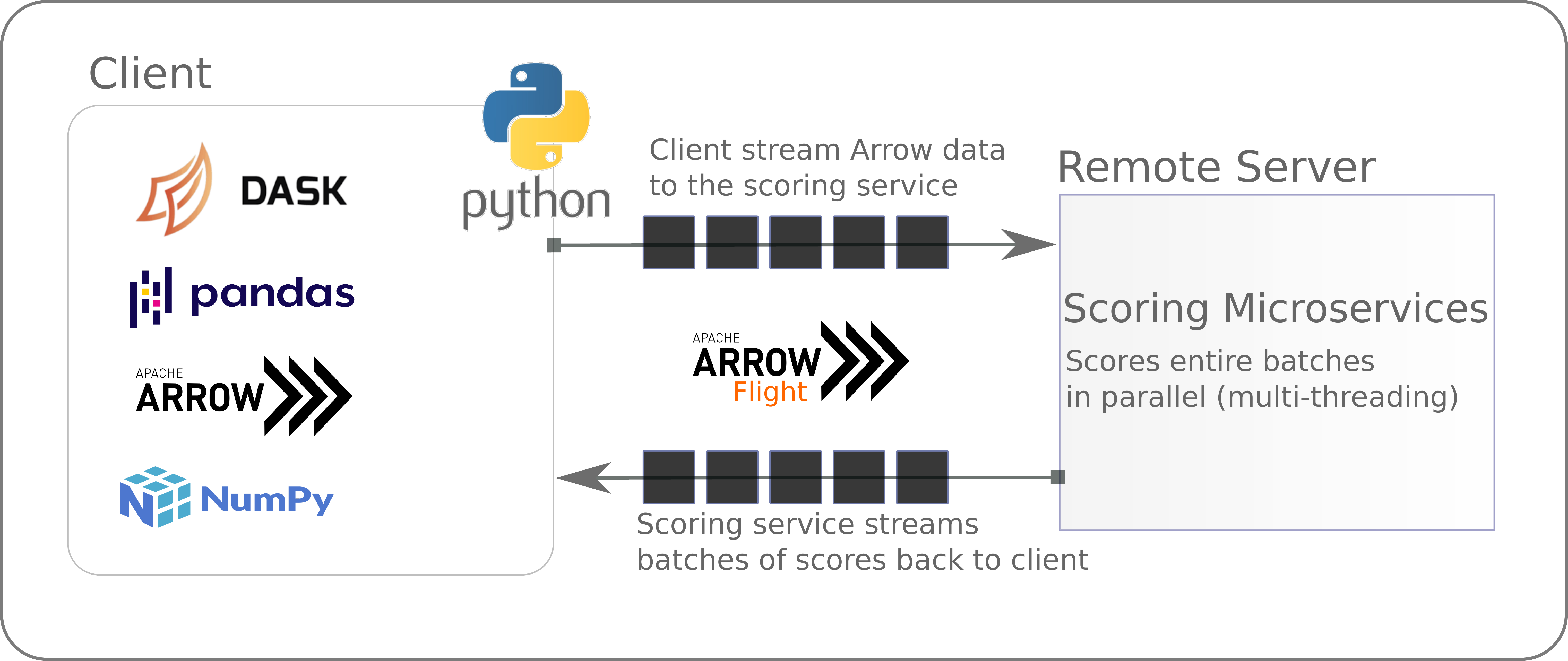}
  \caption{A remote scoring microservice using Arrow data batches and communicating via Arrow Flight.}
  \label{fig:xgbatch}
\end{figure}

ML model deployment generally consists of two phases. First, models are trained and validated with existing datasets to uncover pattern and correlations within the data. Then, the best performing trained model is applied to new datasets, to perform various tasks, such as predicting the probability scores in the case of classification problems, or estimating averages in the case of regression problems~\cite{scoring}. In production environments, ML based applications usually have separate deployment methods for real time model needs (e.g., an API or gRPC service, etc.) vs batch scoring (e.g., some form of Spark or Dask based solution)~\cite{xgbatch}. Real time use cases need low latency for processing millions of records each row at a time, while batch processes need to take advantage of modern hardware features like multi-cores, vectorization, accelerators (GPUs/FPGAs) and high throughput interconnects on cluster environments to process and transfer the large amount of data quickly. XGBATCH as shown in Figure~\ref{fig:xgbatch} uses Apache Arrow's Flight framework (which is built on the gRPC protocol under the hood) to stream batches of data to the scoring service, which in-turn scores it as a batch, and finally streams the batch back to the client. Using Flight ensures low latency for real time use cases, as well as an efficient method for scoring large batches of data.    
\subsubsection{FlightGrid/PyGrid - AI Models Training~\cite{FlightGrid}}
PyGrid is a peer-to-peer platform for secure, privacy-preserving and decentralized data science and analytics. Data owners and scientists can collectively train AI models using the PySyft framework. In PyGrid data-centric federated learning (FL) use cases, a lot of data movement between domain and workers network is involved. In a FlightGrid implementation for a simple network using mnist dataset with batch size 1024 and pre-trained model with Arrow data format on Arrow Flight nodes shows more than 5x speedup with the same accuracy as compared to regular grid data.   
\subsubsection{The Mesh for Data platform - Arrow/Flight module~\cite{AFM}}
The Mesh for Data is a cloud-native platform to control the data usage within an organization premises. It provides a platform to unify data access, governance and orchestration, enabling business agility while securing enterprise data. The arrow-flight-module (AFM) for The Mesh for Data brings enforcement of data governance policies to the world of Apache Arrow Flight for fast and efficient data movement and analytics within applications to consume tabular data from the data sources. Currently, AFM provides support for a couple of different file-systems, formats and queries for Arrow datasets.
\label{subsec:Future}
\section{Future Outlook}
Currently Flight service operations rely only on TCP data transport layer. Using gRPC to coordinate get and put transfers on protocols other than TCP like RDMA have huge potential to speed up bulk data transfers over the networks with RDMA support. As shown in~\cite{mainliningDBs} for some typical databases, the data export speed to some external applications while the majority of the blocks is in a frozen state can utilize up to 80\% of total available network bandwidth. This result suggests that bypassing the network stack for Arrow bulk data transfers via RDMA can easily saturate high bandwidths networks. 
The Flight SQL proposal~\cite{FlightSQL} which is being implemented paves the way for client-server databases to directly communicate with SQL-like semantics. This feature will enable browsing database metadata and execution of queries while transferring data streams with Arrow Flight. 

In the context of distributed systems,
many different distributed columnar databases and query engines also propose to integrate an Arrow Flight layer support for data export to external applications/transfer bulk data in Arrow supported frameworks. In addition, many distributed AI/ML training and inference workloads~\cite{tensorio, xgbatch} are also being equipped with Arrow Flight functionality.





\section{Limitations of this work}
\label{subsec:Limitations}
This work is an early of what is possible with Arrow format and Arrow Flight as an Arrow data transfer, querying and microservice context. Arrow APIs including Arrow Flight are under heavily development process for both new features addition and performance improvements. At the same time all the projects discussed in this article are also under development. We believe in coming months these projects will be matured enough to be integrated into existing frameworks for both better performance and scalability.      
\section{Conclusion}
\label{subsec:Conclusion}
Apache Arrow is a columnar in-memory data format which provides cross-language support for data analytic applications and frameworks. It enables fast data movement within big data frameworks eco-system by avoiding (de)-serialization overheads. Arrow Flight is gRPC based framework which provides high speed data communication services for Arrow data transfer over the networks. In this article, we demonstrated and benchmarked a couple of Arrow Flight use cases. For bulk Arrow data transfer we benchmarked the throughput on both local and remote hosts with varying batch sizes on a 7000 MB/s inter-node bandwidth. The maximum 1650 MB/s throughput achieved for DoPut() while DoGet() achieves upto 2000 MB/s throughput with upto 16 streams in parallel on remote hosts. On local machine Arrow Flight achieves upto 10K MB/s throughput. In genomics pipeline, the distributed regions specific chromosomes sorting of ArrowSAM~\cite{ArrowSAM} data achieves upto 500 MB/s throughput. Note that in this particular scenario, all nodes are connected through Flight endpoints and sending/receiving Arrow RecordBatch streams at the same time in parallel. We also included the results of DB-X bulk data export speeds of different (client-side RDMA, Arrow Flight, vectorized wire protocol and row-based PostgreSQL wire protocol) protocols, where Flight protocol uses nearly 80\% of total available bandwidth in case all blocks are frozen. By comparing the results of different data querying APIs like ODBC, turbodbc and Flight on a Dremio client also shows a significant performance/data transfer time improvement when accessing/querying somehow big size datasets. Arrow Flight based implementation on Dremio performs 20x and 30x better as compared to turbodbc and ODBC connections respectively. Rust based Datafusion Flight API also provides client-server implementation for SQL querying on CSV and Parquet data over Flight. Moreover, we also analysed some microservices uses cases like Apache Spark and TensorFlow clients to put/get Arrow data streams in parallel where Flight can be used as a fast Arrow data transfer layer to speedup the analytical processes on batches. Reading multiple Flight endpoints in parallel as Spark partitions in a multi-node cluster as compared to existing serial JDBC approach in Spark improves the performance by many folds. Batch scoring/processing and remote ML models training/testing on single as well as multi-node cluster environments on Arrow data through Flight has potential to improve the performance of existing algorithms by an orders-of-magnitude.                         

\printbibliography

\end{document}